\documentclass[fleqn,usenatbib]{mnras}

\usepackage{newtxtext,newtxmath}

\usepackage[T1]{fontenc}
\usepackage{ae,aecompl}

\usepackage{graphicx}	
\usepackage{amsmath}	
\usepackage{amssymb}	

\usepackage{listings}
\usepackage{multirow}
\usepackage{units}
\usepackage{ulem}

\newcommand{\tabitem}{\protect~~\llap{\textbullet}~~}

\title[Relations for p-Mode Sensitivities]{Empirical Relations for the Sensitivities of Solar-like Oscillations to Magnetic Perturbations}

\author[R. Kiefer \& A.-M. Broomhall]{
Ren\'e Kiefer\thanks{E-mail: r.kiefer@warwick.ac.uk},
Anne-Marie Broomhall
\\
Centre for Fusion, Space, and Astrophysics, Department of Physics, University of Warwick, Coventry, CV4 7AL, United Kingdom
}

\date{Accepted XXX. Received YYY; in original form ZZZ}

\pubyear{2020}

\begin{document}
\label{firstpage}
\pagerange{\pageref{firstpage}--\pageref{lastpage}}
\maketitle

\begin{abstract}
Oscillation mode frequencies of stars are typically treated as static for a given stellar model. However, in reality they can be perturbed by time varying sources such as magnetic fields and flows. We calculate the sensitivities of radial p-mode oscillations of a set of models for masses between \unit[0.7--3.0]{M$_{\odot}$} from the main sequence to the early asymptotic giant branch. We fit these mode sensitivities with polynomials in fundamental stellar parameters for six stages of stellar evolution. We find that the best-fitting relations differ from those proposed in the literature and change between stages of stellar evolution. Together with a measure of the strength of the perturbation, e.g., of the level of magnetic activity, the presented relations can be used for assessing whether a star's observed oscillation frequencies are likely to be close to the unperturbed ground state or whether they should be adjusted. 
\end{abstract}



\section{Introduction}\label{sec:1}
Owing to the revolution that the high-quality space-photometry data of \textit{Kepler} \citep{Borucki2010, Koch2010}, CoRoT \citep{Auvergne2009, Baglin2006}, and TESS \citep{Ricker2014} have brought about, many thousands of stars now have measured oscillation frequencies. These stars cover very different stages of evolution from the main sequence \citep[e.g.,][]{Appourchaux2012, Lund2017}, to subgiants \citep{Appourchaux2012}, red giants \citep[e.g.,][]{Corsaro2015, Kallinger2019}, core helium burning and asymptotic giant branch (AGB) stars \citep{Kallinger2019}, and even white dwarfs \citep{Hermes2017}. Their oscillation frequencies can be subject to perturbations due to intrinsic time varying sources, for example magnetic activity: it is well known that for the Sun p-mode frequencies of low-degree modes at the frequency of maximum oscillation amplitude $\nu_{\text{max}}$ vary by about \unit[0.4]{$\mu$Hz} over the course of the solar activity cycle \citep{Woodard1985, Elsworth1990, Libbrecht1990, Jimenez-Reyes1998}. Similar behaviour has been observed for other main-sequence stars \citep[e.g.,][]{Garcia2010, Kiefer2017, Santos2018}.

Stellar activity appears to persist all the way through a star's evolution from the main sequence to the AGB: activity cycles have been detected on dozens of main-sequence stars \citep[e.g.,][]{Baliunas1995}; \cite{Auriere2015} measured magnetic fields at the surface of K and G giants; starspots have been detected on various types of stars \citep[see references in][]{Berdyugina2005}. Whether activity is generated by global cyclic dynamos or by local turbulent mechanisms is, in first approximation, not relevant for the perturbation of mode frequencies.

Several frameworks for understanding how various perturbations affect oscillation mode parameters have been developed, mostly with the Sun as the benchmark: \cite{Dziembowski2004} used a variational approach to quantify the effect of the change of dynamical quantities over the solar cycle on mode frequencies. Coupling of normal-modes within the framework of quasi-degenerate perturbation theory was investigated by \cite{Roth2002a} to calculate the effect of large-scale flows on p modes, by \cite{Schad2013a} to size the meridional circulation, and by \cite{Kiefer2017b} and \cite{Kiefer2018} to assess the effect of global toroidal magnetic fields on p-mode frequencies and eigenfunctions. \cite{Hanasoge2017} also exploits normal mode coupling, but with first-Born perturbation theory, to gauge the impact of Lorentz stresses on p-mode oscillations.

Recently, \cite{Howe2020} found that the solar $\nu_{\text{max}}$ varies by as much as \unit[25]{$\mu$Hz} over the Sun's activity cycle. In an asteroseismic context, it may thus be important to account for activity related variations of $\nu_{\text{max}}$ and perturbations of measured mode frequencies when they are used to model the stellar structure or to infer global stellar parameters such as radius and mass \citep[see also][]{Kiefer2019, PerezHernandez2019}. The magnitude of the frequency shifts can be estimated by the product of a factor accounting for the sensitivity of the oscillations to a perturbation and a factor for the strength of the perturbation, as we will detail in Section~\ref{sec:3}. In their effort to estimate the number of stars with detectable activity related p-mode frequency shifts in TESS data, \cite{Kiefer2019} derived the mode sensitivity $S$ to be proportional to $R M^{-1}\nu_{\text{max}}^{-1}$, where $R$ is the stellar radius and $M$ is stellar mass. This is very similar to the modal sensitivity contribution which \cite{Karoff2009} gave using the relation of \cite{Metcalfe2007}, which was $S\propto R^{2.5} L^{0.25} M^{-2}$, where $L$ is stellar luminosity. The scaling suggested by \cite{Chaplin2007} did not incorporate a factor for mode sensitivity, but simply assumes linear scaling of mode frequency shifts with the level of activity.

The exponents in these relations are fixed and it is thus expected that they cannot capture the variation of the mode sensitivity over the entire evolution from the main sequence to the early AGB. In this paper, we first describe the stellar evolutionary models we use to calculate the mode sensitivity in Section~\ref{sec:2}. The derivation of the mode sensitivity factor follows in Section~\ref{sec:3}. We then fit these sensitivities for six stages of stellar evolution with several polynomials in stellar fundamental parameters, as described in Section~\ref{sec:4}. The results are discussed in Section~\ref{sec:5} and the conclusions we draw from these results follow in Section~\ref{sec:6}.


\section{Models}\label{sec:2}
We computed stellar models using Modules for Experiments in Stellar Astrophysics (MESA)\footnote{\hyperlink{http://mesa.sourceforge.net/}{http://mesa.sourceforge.net/}}, revision 12115 \citep{Paxton2011, Paxton2013, Paxton2015, Paxton2018, Paxton2019}, for fifteen initial stellar masses and evolved them to the early AGB. We used the MESA test suite inlist for the evolution of a \unit[1]{M$_{\odot}$} star which is stored in the MESA installation's folder \nolinebreak{\path{'\star\test_suite\1M_pre_ms_to_wd
'}}, adjusted the initial mass and, in order to have a better resolution of the main-sequence and the core helium burning phases, the maximum time between two models. The initial masses and their respective maximum time step values are given in Table~\ref{table:1}. We terminated the evolution at the thermally pulsing (TP)-AGB, but as we are only considering stellar models with $\nu_{\text{max}}$\,>\,\unit[1]{$\mu$Hz} (see Table~\ref{table:2}), the more evolved models on the AGB are not included in our study. A sample MESA inlist is given in Appendix~\ref{sec:app:1.1} but will also be made available online. 

The models use the solar metal mixture of \cite{Grevesse1998} with a helium abundance of $Y_{\odot}=0.248$ and a metallicity of $Z_{\odot}=0.017$. The atmospheric model uses the MESA default, a standard Eddington grey atmosphere. The opacities are those of \cite{Iglesias1993, Iglesias1996} and the MESA option for C/O enhancement during and after helium burning is enabled with a base metallicity of 0.02.

The adiabatic radial p-mode eigenfunctions and eigenfrequencies of all models were computed with the oscillation code GYRE\footnote{\hyperlink{https://bitbucket.org/rhdtownsend/gyre/}{https://bitbucket.org/rhdtownsend/gyre/}} \citep{Townsend2013, Townsend2018}. A sample GYRE inlist is given in Appendix~\ref{sec:app:1.2} but will also be made available online.

\begin{figure}
	\includegraphics[width=1\columnwidth]{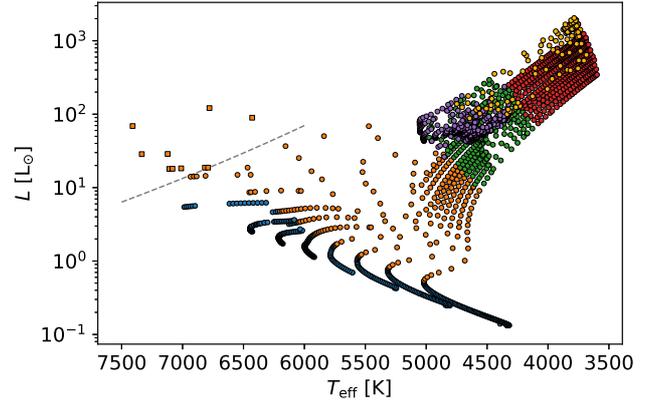}\caption{Hertzsprung-Russell diagram of the stellar models used in this study. Colours encode stellar evolutionary stages as defined in Table~\ref{table:2}: main-sequence stars are blue, subgiants and lower red giant branch stars are orange, upper RGB models are green, tip RGB models are red, core helium burning models are purple, and AGB models are amber. The red edge of the instability strip is indicated by the dashed grey line.}\label{fig:1}
\end{figure}

Figure~\ref{fig:1} shows a Hertzsprung-Russell diagram of the models we selected based on the requirements that they have a convection zone with a top boundary above $\unit[0.99]{R}$, they have reached the main sequence, and their expected $\nu_{\text{max}}$ is greater than \unit[1]{$\mu$Hz}. The requirement on the outer convection zone boundary ensures that p modes are detectable at the surface and are not suppressed by a thick outer radiative zone. We then separated the remaining models into six evolutionary stages based on the conditions listed in the second column of Table~\ref{table:2}. The third column gives the final number of models in each stage. The partitioning into main sequence, subgiant and the red giant branch (RGB), core He burning, and AGB is founded on different stages of energy production and core composition. The subdivision of the RGB into lower RGB, upper RGB, and tip RGB models was guided by the phenomenology of the mode sensitivities. In particular, the separation of upper and tip RGB models improved the quality of the polynomial fits, as will be described in Section~\ref{sec:4}. Ten subgiant models are bluewards of the red edge of the instability strip defined by \cite{Chaplin2011} as
\begin{align}
    T_{\text{red}} = \unit[8907]{K}\cdot L^{-0.093},\label{sec2:eq1}
\end{align}
where the luminosity $L$ is in solar units. These models have square symbols in Figure~\ref{fig:1} and in all following plots.

\begin{table}
	\caption{Initial masses and maximum time steps between two models.}\label{table:1}
\begin{center}
	\begin{tabular}{cc}
		\hline
		Initial mass $[\unit{M_{\odot}}]$ & Max. time step $[\unit{yr}]$ \\ \hline
		   0.7,\,0.8,\,0.9,\,1.0    &        5$\cdot 10^8$        \\ \hline
		         1.1,\,1.2          &        1$\cdot 10^8$        \\ \hline
		      1.3,\,1.5,\,1.7       &        5$\cdot 10^7$        \\ \hline
		       1.8,\,2.0,\,2.2        &        2$\cdot 10^7$        \\ \hline
		      2.5,\,2.8,\,3.0       &        1$\cdot 10^7$        \\ \hline
	\end{tabular}
\end{center}
\end{table}


\begin{table*}
	\caption{Separation of evolutionary stages, number of models in each stage, relation in Table~\ref{table:3} which gives the lowest mean absolute error between fit and model sensitivities for each stage, and letter symbol that is used to identify this stage.}
	\label{table:2}
	\begin{center}
		\begin{tabular}{|c|l|c|c|c}
			\hline
			Stage     & \multicolumn{1}{c}{Conditions}                                      & \# models & Best-fitting relation & ID \\ \hline
			Main sequence & \tabitem  Centre hydrogen content $>10^{-5}$                        &    383    &            5          &             A             \\
			& \tabitem$\nu_{\text{max}}>$\unit[1000]{$\mu$Hz}                     &           &                       &                           \\
			& \tabitem$T_{\text{eff}}<$\unit[7000]{K}                             &           &                       &                           \\
			& \tabitem Contraction is finished $L_{\text{tot}}-L_{\text{H}}<0.01$ &           &                       &                           \\ \hline
			Subgiants \&  & \tabitem Centre hydrogen content $<10^{-5}$                         &    230    &           6           &             B             \\
			lower RGB   & \tabitem $\nu_{\text{max}}>$\unit[100]{$\mu$Hz}                     &           &                       &                           \\
			& \tabitem $T_{\text{eff}}<$\unit[7500]{K}                            &           &                       &                           \\ \hline
			Upper RGB   & \tabitem Centre hydrogen content $<10^{-5}$                         &    257    &           6           &             C             \\
			& \tabitem \unit[10]{$\mu$Hz}< $\nu_{\text{max}}$<\unit[100]{$\mu$Hz} &           &                       &                           \\
			& \tabitem Helium core not ignited $L_{\text{He}}<0.1L_{\text{H}}$    &           &                       &                           \\ \hline
			Tip RGB    & \tabitem Centre hydrogen content $<10^{-5}$                         &    382    &           4           &             D             \\
			& \tabitem \unit[1]{$\mu$Hz}< $\nu_{\text{max}}$<\unit[10]{$\mu$Hz}   &           &                       &                           \\
			& \tabitem Helium core not ignited $L_{\text{He}}<0.1L_{\text{H}}$    &           &                       &                           \\ \hline
			Core He burning   & \tabitem Centre hydrogen content $<10^{-5}$                         &    514    &           5           &             E             \\
			& \tabitem Centre helium content $>10^{-5}$                           &           &                       &                           \\
			& \tabitem $\nu_{\text{max}}>$\unit[10]{$\mu$Hz}                      &           &                       &                           \\
			& \tabitem Helium burning  $L_{\text{He}}> 0.1L_{\text{H}}$           &           &                       &                           \\ \hline
			AGB      & \tabitem Centre hydrogen content $<10^{-5}$                         &    175    &           7           &             F             \\
			& \tabitem Centre helium content $<10^{-5}$                           &           &                       &                           \\
			& \tabitem $\nu_{\text{max}}>$\unit[1]{$\mu$Hz}                       &           &                       &                           \\ \hline
			All      & \tabitem Top convection zone boundary $\ge$\unit[0.99]{$R$}         &     1941      &                       &                           \\ \hline
		\end{tabular} 
	\end{center}
{\raggedright \textbf{Notes.} $L_{\text{tot}}$ is the total luminosity of the star including all sources. $L_{\text{H}}$ and $L_{\text{He}}$ are the contributions to the total luminosity from hydrogen and helium burning processes, respectively \par}
\end{table*}

\section{Mode Sensitivity}\label{sec:3}
We start with the variational expression for frequency shifts derived by \cite{Dziembowski2004} (compare to their equation 18) and quoted by \cite{Metcalfe2007} in their equation 1:
\begin{align}
	\delta\nu_i = \frac{\int dV\, \mathcal{K}_i\mathcal{S}_i}{2I_i\nu_i},\label{sec3:eq1}
\end{align}
where $\mathcal{K}_i$ is the kernel function relating perturbations caused by a source function $\mathcal{S}_i$ to the shifts $\delta\nu_i$, $I_i$ is the mode inertia, and $\nu_i$ is the frequency of mode $i$ in $\unit{\mu Hz}$. The integral is calculated over the entire stellar volume $V$. In the following, we will ignore the contribution of magnetic activity to Eq.~(\ref{sec3:eq1}) and assume that the resulting frequency shift is linear in the level of activity: $\delta\nu= S\cdot A$, where $S$ is the sensitivity of a mode to a perturbation and $A$ is a measure of the level of activity. 

Just as in \cite{Metcalfe2007}, we shall assume that the dominant term in the kernel of p modes is proportional to $|\text{div}\boldsymbol{\xi}|^2$, where $\boldsymbol{\xi}$ is the displacement eigenfunction. We consider only radial modes, which considerably simplifies the calculations. With this, the kernel function is
\begin{align}
\mathcal{K} \propto \left|\frac{2}{r}\xi_r(r) + \frac{\partial \xi_r(r)}{\partial r}\right|^2,\label{sec3:eq2}
\end{align}
where $\xi_r(r)$ is the radial displacement eigenfunction at radial position $r$. 

We assume the perturbation is located within one pressure scale height $H_{\text{p}}$ below the photosphere with uniform and normalized strength 
\begin{align}
\mathcal{S} &= \Theta(H_{\text{p}}-d)=\begin{cases}
1, & \text{if}\ H_{\text{p}}-d>0 \\
0, & \text{otherwise.}
\end{cases}\label{sec3:eq3}
\end{align}
Here, $\Theta$ is the Heaviside step function and $d$ is the depth below the photospheric radius of the model. Thus, in this article we make no assumption about the level of magnetic activity affecting the modes or how this evolves over the stellar lifetime. 

With this definition of the source function, we explicitly concentrate on (near-)surface magnetic activity affecting mode frequencies. From the Sun, it is known that the activity related frequency shifts largely originate from the near-surface regions \citep[e.g.,][]{Broomhall2017}. Similar behaviour has been observed on other solar-like oscillators by, e.g., \cite{Salabert2011, Salabert2016a} and \cite{Kiefer2017}. Deeper located magnetic fields or core magnetic fields are likely to affect the temporal stability of mode frequencies less \citep[e.g.,][]{Gough1990, Kiefer2018}. This is because such magnetic fields have a much smaller effect on p modes than near-surface fields due to the drop in sensitivity. Further, core fields are assumed to be stable over very long time scales, meaning they are unlikely to bring about a mode frequency variation on observable time scales. The deeper a magnetic field is located, the stronger it has to be to achieve the same level of frequency shift. We will discuss this further in Section~\ref{sec:52}.

In Eq.~(\ref{sec3:eq1}), the mode inertia $I$ is calculated as 
\begin{align}
	I = \int dV \rho |\boldsymbol{\xi}|^2,\label{sec3:eq4}
\end{align}
where the integral extends over the stellar volume $V$, $\rho$ is density, and the displacement eigenfunction $\boldsymbol{\xi}$ can again be replaced by $\xi_r(r)$, as we consider only radial modes.

For simplicity, we investigate only the oscillation mode $i$ closest to $\nu_{\text{max}}$, which is calculated for each model with the scaling relation of \cite{Kjeldsen1995}:
\begin{align}
\nu_{\text{max}} = \nu_{\text{max}\odot}\cdot \left(\frac{M}{M_{\odot}}\right)\left(\frac{R}{R_{\odot}}\right)^{-2}\left(\frac{T_{\text{eff}}}{T_{\text{eff}\odot}}\right)^{-\frac{1}{2}}\,,\label{sec3:eq5}
\end{align}
with the calibrated solar reference value $\nu_{\text{max}\odot}=$~\unit[3104]{$\mu$Hz} of \cite{Mosser2013}, the solar values of mass $M_{\odot}=\unit[1.9892\cdot 10^{33}]{g}$, radius $R_{\odot}=\unit[6.9598\cdot 10^{10}]{cm}$, and effective temperature $T_{\text{eff}\odot}=\unit[5777]{K}$ as used in the MESA code. $R$ and $M$ are the model's photospheric radius and mass, respectively. 

With the above definitions Eqs.~(\ref{sec3:eq2}), (\ref{sec3:eq3}), (\ref{sec3:eq4}) and the stated assumptions, we calculate the right-hand side of Eq.~(\ref{sec3:eq1}) for the set of stellar models described in Section~\ref{sec:2} and call this the mode sensitivity $S$ of a star:
\begin{align}
	S = \frac{\displaystyle\int_{H_{\text{p}}}^{R} dV\, \mathcal{K}_{\text{i,max}}}{\displaystyle 2I_{\text{i,max}}\nu_{\text{i,max}}},\label{sec3:eq5.2}
\end{align}
where the index 'i,max' indicates the mode closest to $\nu_{\text{max}}$. The unit of $S$ is $\unit{cm\,g^{-1} \mu Hz^{-1}}$.

To better illustrate how the observed behaviour of the mode sensitivities comes about, we show the three main quantities that make up the sensitivities, separately:
\subsection{Pressure scale height}\label{sec3.1}
The vertical distance over which hydrostatic pressure increases by a factor of $e$ in a stratified atmosphere can be written as
\begin{align}
    H_{\text{p}} \propto \frac{P}{\rho g} \propto L^{0.25} R^{1.5} M^{-1}\label{sec3:eq6}
\end{align}
where it was taken that $\rho \propto M R^{-3}$ and $g\propto MR^{-2}$, where $g$ is surface gravity. It was assumed that the stellar photosphere follows the equation-of-state of an ideal gas $P\propto\rho T$, where $P$ is pressure and $T$ is temperature. Further, the Stefan-Boltzmann law was used to express temperature as $T\propto L^{0.25}R^{-0.5}$.

\begin{figure*}
	\includegraphics[width=\linewidth]{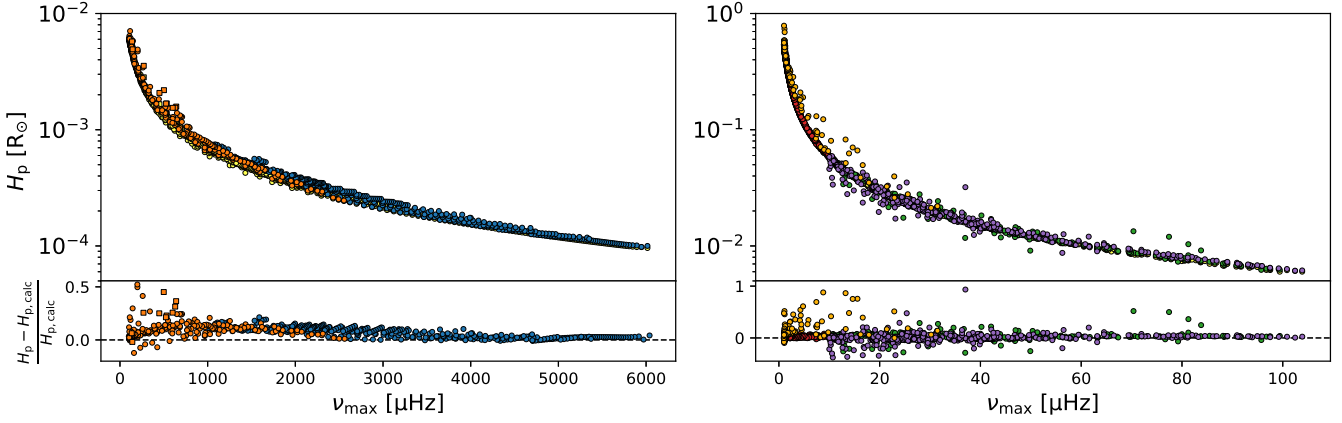}
	\caption{Pressure scale height $H_{\text{p}}$ calculated from the MESA models as a function of $\nu_{\text{max}}$ for the six stages of stellar evolution as indicated in Table~\ref{table:2} with colours as in Fig.~\ref{fig:1}. Left panel comprises the first two stages, right panel the other four stages. Pressure scale heights calculated with Eq.~(\ref{sec3:eq6}) are shown in yellow, but are largely covered by the models' data points. The bottom section of both panels shows the normalized difference between the calculated and the models' scale height.}\label{fig:2}
\end{figure*}

Figure~\ref{fig:2} shows the pressure scale heights from the stellar models with the same colour code for the six stages of stellar evolution as described in Fig.~\ref{fig:1} and detailed in Table~\ref{table:2}. It is calculated for each model as the depth of that grid point at which the pressure is closest to $e\cdot P(\text{photosphere})$. The left panel shows the first two stages and the right panel comprises the last four stages in order to increase clarity as $\nu_{\text{max}}$ is the ordinate. The scale heights $H_{\text{p, calc}}$ calculated using Eq.~(\ref{sec3:eq6}) are shown in yellow data points. The values are scaled such that Eq.~(\ref{sec3:eq6}) yields a value of \unit[150]{km} for the Sun.

As the models are in hydrostatic equilibrium, the models' pressure scale heights and the value obtained from Eq.~(\ref{sec3:eq6}) are within a few percent: the data points largely overlap. The bottom sections of the panels in Fig.~\ref{fig:2} show the normalized difference between the two sets of values. The ideal gas assumption and that of a perfect black body will obviously affect the degree of accuracy of the rather simplistic Eq.~(\ref{sec3:eq6}) through stellar evolution where photospheric conditions highly vary. The largest relative systematic deviation between the models' scale heights and those from Eq.~(\ref{sec3:eq6}) are found during the subgiant stage. However, being within a few tens of percent of the model values, Eq.~(\ref{sec3:eq6}) is a good approximation of stellar photospheric pressure scale heights.
\subsection{Integral over kernel function}\label{sec3.2}
The integral extends over the stellar volume of the outer pressure scale height from the photospheric radius inwards. We assume the source function $\mathcal{S}$ to be unitless. As the unit of the eigenfunctions is $\unit{cm}$, the unit of the numerator in Eq.~(\ref{sec3:eq1}) is $\unit{cm^3}$. 

\begin{figure*}
	\includegraphics[width=\linewidth]{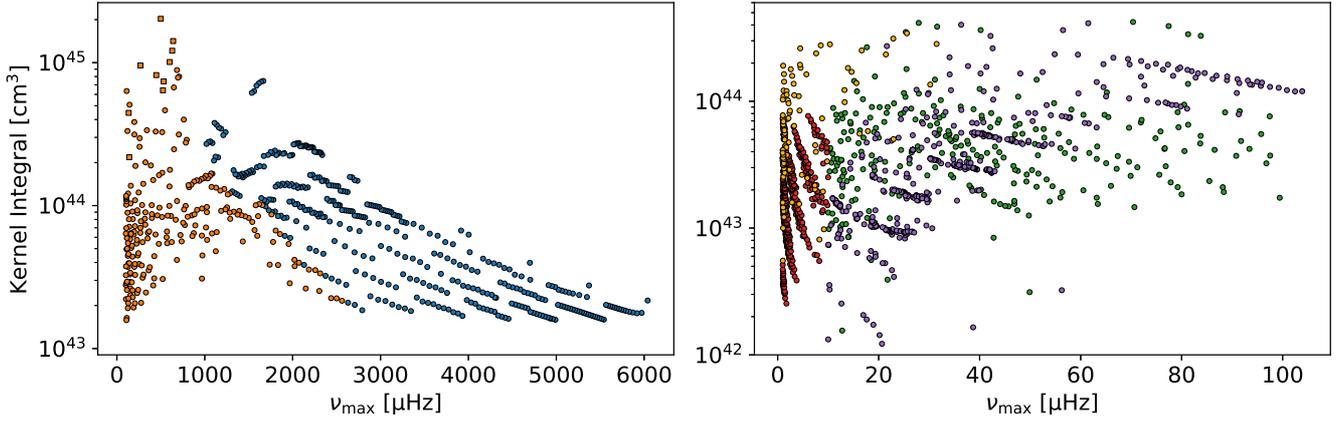}
	\caption{Numerator of Eq.~(\ref{sec3:eq1}) with model scale heights as shown in Fig.~\ref{fig:2} over $\nu_{\text{max}}$ for the six stages of stellar evolution as indicated in Table~\ref{table:2} with colours as in Fig.~\ref{fig:1}. Left panel comprises the first two stages, right panel the other four stages.}\label{fig:3}
\end{figure*}

The integral values of all investigated models are presented in Figure~\ref{fig:3} with colours and separation into two panels as in Fig.~\ref{fig:2}. On the main sequence, models around $\nu_{\text{max}}\approx\unit[2000]{\mu Hz}$ have the largest integral values. Higher mass stars have higher values than lower mass stars at same $\nu_{\text{max}}$. For the subgiant and lower RGB stars there is a spike in integral values for masses $\unit[1.5-2]{M_{\odot}}$ around $\nu_{\text{max}}\approx\unit[600]{\mu Hz}$. This increase is due to a surge in the pressure scale height for these stars, thus more of the stellar radius and the kernel function are integrated over. The ridges correspond to different radial orders of the mode closest to $\nu_{\text{max}}$. For the models in the tip RGB phase (red data points), the five ridges correspond to radial orders $n=3-7$, counting from lowest to highest $\nu_{\text{max}}$. To increase clarity of the figure, a few core He burning models with integral values $<10^{42}$ were cut. 

\begin{figure*}
	\includegraphics[width=\linewidth]{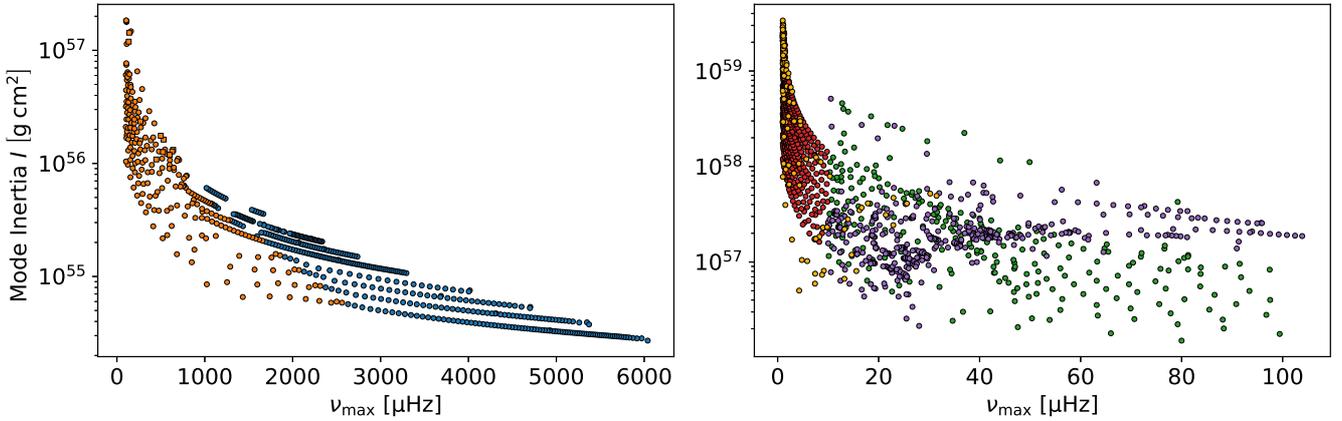}
	\caption{Same as Fig.~\ref{fig:3} but for mode inertia as calculated by use of Eq.~(\ref{sec3:eq4}).}\label{fig:4}
\end{figure*}
\subsection{Mode inertia}\label{sec3.3}
The integral in Eq.~(\ref{sec3:eq4}) extends over the entire stellar volume and amounts to a weighting of the stellar density with the eigenfunctions being the weights. Mode inertia is of the unit $\unit{cm^2\,g}$. Figure~\ref{fig:4} shows the mode inertias with colours and separation into two panels as in Fig.~\ref{fig:2}. The ridges correspond to different mass values, with lower masses having smaller inertias. As the radial modes that are investigated here propagate throughout the star, mode inertias tend to be larger for higher mass stars. This also explains the steep increase with lower $\nu_{\text{max}}$.


\section{Fits of Polynomials of Stellar Parameters}\label{sec:4}
\begin{figure*}
	\includegraphics[width=\linewidth]{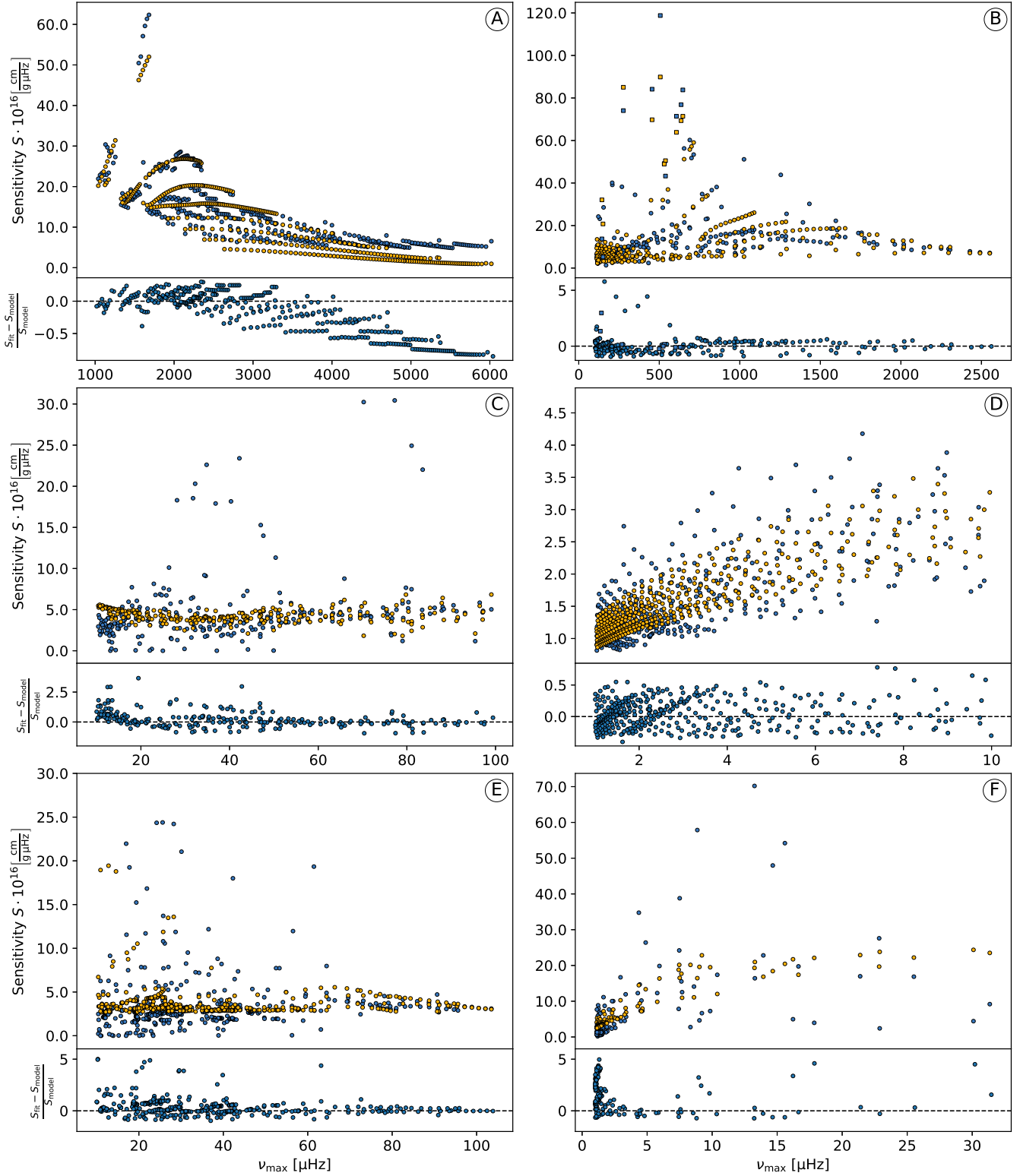}\caption{Fits of polynomials in fundamental stellar parameters to the calculated sensitivity factor $S$. Blue data points are model sensitivities and amber circles are the results of the best-fitting relation presented in Table~\ref{table:4}. Panels show  the six evolutionary stages as laid out in Table~\ref{table:2}. Note that the ordinates	and abscissae of the panels have differing value domains. The bottom section of each panel shows the normalized residuals between fitted and model sensitivities.}\label{fig:5}
\end{figure*}

The mode sensitivity value $S$ of the mode eigenfunction closest to each model's estimated $\nu_{\text{max}}$ was calculated as described in Section~\ref{sec:3}. For the ease of plotting, $S$ was multiplied by a factor of $10^{16}$ before the fits were carried out. Figure~\ref{fig:5} shows the values of $S$ obtained as a function of $\nu_{\text{max}}$ for each evolutionary stage separately with letter labels in the top right of each panel as given for the different evolutionary stages in Table~\ref{table:2}. Various polynomial models, which are described in the second column of Table~\ref{table:3}, were then fitted to $S$ independently for each evolutionary stage. The best-fit parameters obtained from collecting 1000 least-squares fits, each with randomly chosen initial values for the parameters, are given in columns three to eight of Table~\ref{table:3}. The initial values were limited to the  half-open interval $\left[-50,50\right)$. Fits for which the absolute of any exponent was >50 were neglected. The fits were repeated until 1000 stable samples were found. The best fit is chosen as that one with the smallest mean absolute error (MAE) between fit and calculated sensitivities. The last column of Table~\ref{table:3} gives the MAE of the best fit. The mean absolute error is calculated as 
\begin{align}
	\text{MAE} = \frac{\sum_{i=0}^N{\left|S_{\text{model,}i}-S_{\text{fit,}i}\right|}}{N-\text{dof}},\label{sec4:eq1}
\end{align}
where, against the usual definition of the MAE, we subtract the number of degrees of freedom of each fit model, dof, from the number of data points $N$ in order to penalize use of more free parameters. 

The fit models were chosen such as to include global stellar parameters which are readily available either from asteroseismic analyses of space telescope light-curves or from spectroscopic data. Thus, models 1 and 4 only include radius $R$, mass $M$, and frequency of maximum oscillation amplitude $\nu_{\text{max}}$, which can be measured from sufficiently long photometric time series. Effective temperature $T_{\text{eff}}$, luminosity $L$, and surface gravity $\log g$, which enter models 2, 3, 5, 6, and 7 have to be obtained from, e.g., Gaia measurements or other spectroscopic observations of the star in question. Model 7 is a solely spectroscopic relation, should seismic parameters and estimates of mass and radius not be available.

Figure~\ref{fig:5} also shows the fits of relations to the models' sensitivity factors. Blue data points are sensitivities calculated from the MESA/GYRE models and amber circles show the best-fitting relation. Note that the ordinates of all panels are frequency of maximum oscillation amplitude $\nu_{\text{max}}$ with differing value domain between panels, as well as that the abscissae of all panels show the model mode sensitivities $S$ with differing domains between panels. In the interest of improved clarity, the parameters of the best-fitting relation for each evolutionary group are given in Table~\ref{table:4}. 

\section{Discussion}\label{sec:5}
\begin{figure}
	\includegraphics[width=\linewidth]{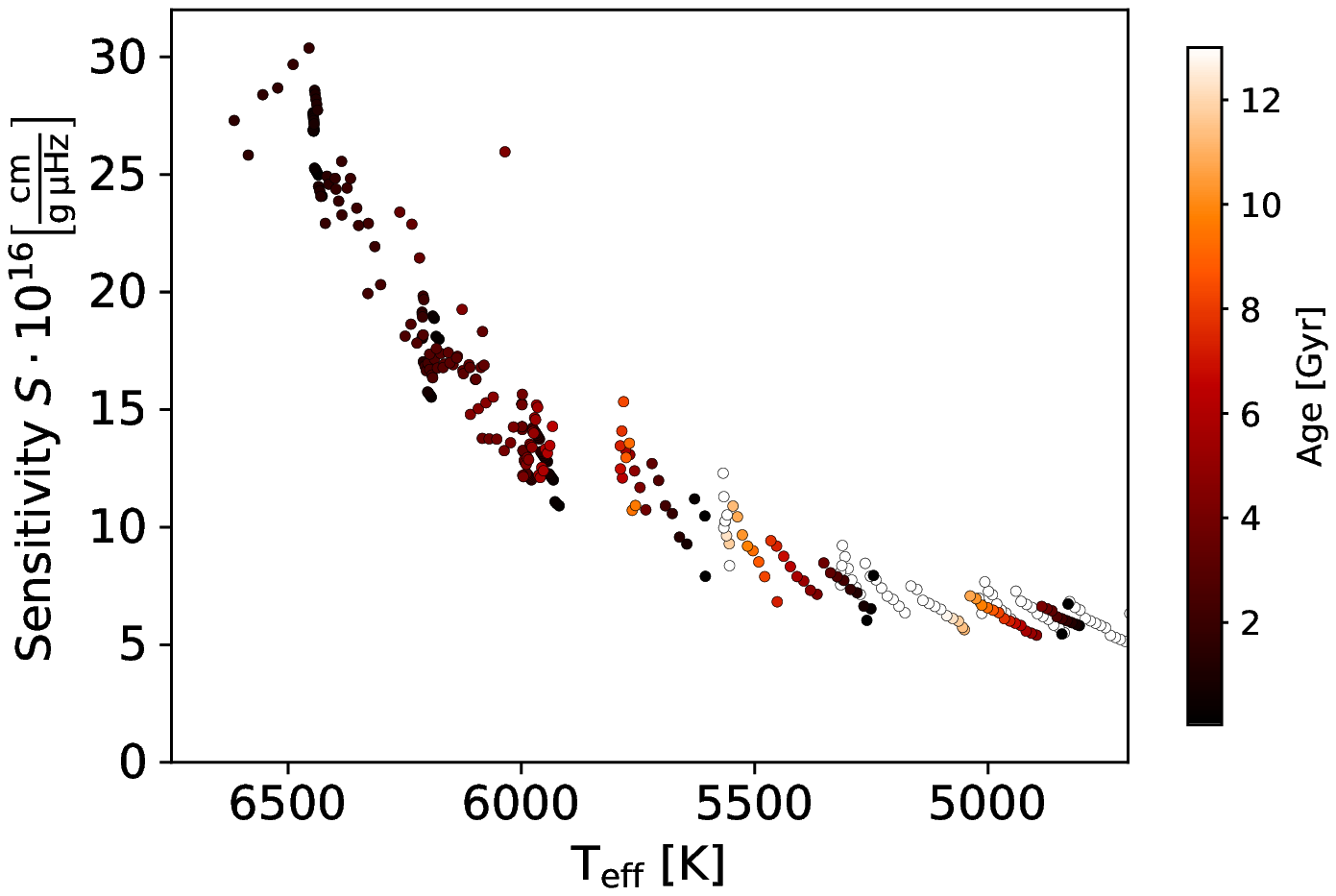}\caption{Mode sensitivities $S$ of the main-sequence models as function of effective temperature $T_{\text{eff}}$. Colors show stellar age with capping of the color bar at $\unit[13]{Gyr}$. }\label{fig:MS_teff}
\end{figure}

The mode sensitivities for main-sequence stars (panel A of Fig.~\ref{fig:5}) are largely determined by the kernel functions. This is because their mode inertias do not change strongly through this stage for each stellar mass. The largest sensitivities are found for stars more massive than the Sun. This agrees with the findings of \cite{Santos2019}, who found larger activity related frequency shift values for stars of higher effective temperature. In Figure~\ref{fig:MS_teff}, the main-sequence models' sensitivities are plotted as a function of effective temperature $T_{\text{eff}}$. The range of $T_{\text{eff}}$ was limited to be the same as in panel c) of Figure 3 of \cite{Santos2019} for easier comparison. The color of the data points indicates stellar age, which, again, was capped at the same value as in \cite{Santos2019}, at $\unit[13]{Gyr}$. By comparison of these two figures, it can be seen that the mode sensitivities calculated here follow a very similar trend as the measured frequency shifts.

The bottom parts of the six panels in Figure~\ref{fig:5} show the normalized residuals between the models' sensitivities and the ones predicted by the best-fitting relations. For the main-sequence models, the residuals are typically a few tens of percent, with larger systematic residuals towards higher values of $\nu_{\text{max}}$. Systematically overestimated sensitivities can also be found for models at the low $\nu_{\text{max}}$ end of panel C (upper RGB) and panel F (AGB). While the residuals can reach up to a factor of a few for some models, the overall agreement between the best-fitting relations' sensitivities and the model sensitivities is much better than for the sensitivity scaling relations from the literature, as we will detail later in this section. For purposes of presentation, we clipped the ordinate of the residuals' panels at 5 for stages B, E, and F. A few single outliers have normalized residuals $>5$.

Subgiants (panel B of Fig.~\ref{fig:5}) have similar mode sensitivities to main-sequence stars. \cite{Isaacson2010} found that $\approx 10\%$ of subgiant stars show moderate to strong chromospheric activity. Thus, even though magnetic activity might not be widespread on this stage of stellar evolution, some stars can be expected to show significant activity related frequency shifts. We also note that given, first, the findings of \cite{Balona2019}, who found evidence of starspots on hot stars, and, second, the large sensitivities of the few models in the instability strip, possible activity-related time variations of radial oscillation mode amplitudes and frequencies of, e.g., $\gamma$ Dor, $\delta$ Scuti, and roAp variables might be worth further investigation. Amplitude as well as frequency variability have been detected on these types of variables by several groups including but not limited to \cite{Holdsworth2014, Breger2016} and \cite{Bowman2016}.

Overall, mode sensitivities decrease for upper (panel C) and tip RGB stars (panel D). However, given their low mode frequencies, even an activity cycle with solar-like amplitude can lead to detectable frequency shifts in such stars. This requires the cycle length to be of the order of the order of the observational baseline. For core helium burning stars (panel E) there are numerous models with  sensitivities of the order of the main sequence again, in particular at around $\nu_{\text{max}}\approx\unit[30]{\mu Hz}$. A few AGB models (panel F) have again rather large mode sensitivities. Depending on the strength and temporal scale of magnetic activity of upper RGB, core helium burning and even early AGB stars, frequency shifts may be detectable for these types of stars in the available \textit{Kepler} dataset.

We do not give a physical interpretation to the best-fitting relation within each evolutionary group, but provide them here as convenient short-cuts to estimate the mode sensitivity of individual stars based on a few readily accessible fundamental parameters. As can be seen from Table~\ref{table:3}, the exponents in each model we fitted can differ quite drastically from one evolutionary stage to the next. That being said, the jumps in the exponents for polynomials 2 and 4 are relatively small and should one be interested in using a single relation for all stars with only changing the exponents from stage to stage, then either one of these appears to be a justifiable choice.

It would in general be possible to obtain uncertainties for the exponents of the polynomial relations, e.g., by using a Markov-chain Monte Carlo estimation of the optimal exponents instead of the least-square fit. However, the MAE of each fit reported in Table~\ref{table:3} can be used as a good estimation of the uncertainty if mode sensitivities are calculated.

To assess the impact of the number of models per evolutionary stage on the fits' results and quality, we redid the fits for two scenarios: first, we halved the number of models by taking every second model, effectively reducing the time resolution. With this, the polynomial relations per evolutionary stage that provide the best fit are the same as with the full model set. The exponents however do change, with the largest changes for the tip RGB and core helium burning stages. For these stages, the best-fit relations are now $R^{6.79}\cdot M^{1.16}\cdot L^{-4.37}\cdot \nu_{\text{max}}^{0.60}$ and $R^{-3.81}\cdot M^{-0.54}\cdot L^{2.62}$, respectively. The residuals and the MAE are on a very similar level as with the full model set for all six stages.

In the second scenario we lowered the number of models to 175 in each stage, which is the number of models of the AGB group. We chose these 175 models randomly from the full set. From this we find similar exponents for the best-fitting relations as for the full model set and the same relations that provide these best fits. The largest changes on the exponents occur again for the tip RGB models, for which the best-fitting polynomial is $R^{4.32}\cdot M^{0.53}\cdot L^{-2.77}\cdot \nu_{\text{max}}^{0.57}$ with a MAE of 0.30. For the core helium burning stage, the number of models is reduced by almost a factor of three (from 514 to 175). The best-fitting model now is $R^{-7.78}\cdot M^{-1.06}\cdot L^{5.04}$, which is quite similar to the one from the full model set. The removal of some of the higher sensitivity models in this stages leads to a somewhat smaller MAE of 1.52. Overall, the MAE of the fits and the residuals are very similar to the full model set.

From these two checks we conclude that the best-fitting relations that we present in Table~\ref{table:4} provide reasonable estimates for the sensitivities in each evolutionary stage. The number of models within each evolutionary group and with it the temporal resolution that we chose for our grid of models appears adequate to capture the bulk behaviour of mode sensitivities.

\begin{figure*}
	\includegraphics[width=1.0\linewidth]{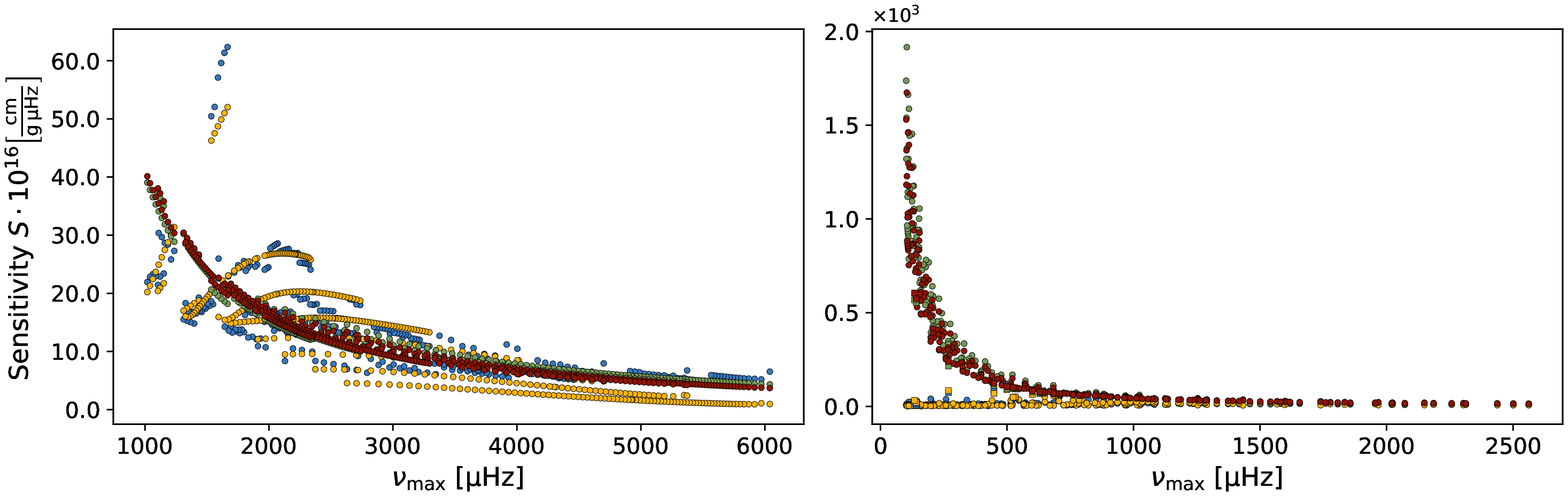}\caption{Comparison of fixed-exponent relations to the best-fitting relation. Blue data points are model sensitivities, amber circles are the results of the best-fitting relation found here. The relations of \protect\cite{Kiefer2019} (green circles) and \protect\cite{Karoff2009}--\protect\cite{Metcalfe2007} (red circles) use the best-fit multiplier based on the main-sequence models only. Left panel is for main-sequence models, right panel for subgiant and lower RGB models.}
		\label{fig:6}
\end{figure*}

In Figure~\ref{fig:6}, the main-sequence (left panel) and subgiant and lower RGB models' (right panel) sensitivity factors of the best fitting relation (amber data points) are compared to those calculated with the relations of \cite{Kiefer2019} (green) and \cite{Karoff2009} based on the derivations of \cite{Metcalfe2007} (red). As mentioned before, these two relations are rather similar -- differing by a factor of merely $\propto\sqrt{T_{\text{eff}}}$ \citep[see][]{Kiefer2019} -- thus, the green data points are largely covered by the red. We limit the comparison to these two evolutionary stages because these relations were not intended for use far off the main sequence. These two relations have fixed exponents and an unknown coefficient which gets lost in their derivation. As this coefficient has to be the same for all evolutionary stages, we obtained it by fitting the relations to the sensitivities of the main-sequence and subgiant and lower RGB models simultaneously. This gave a coefficient of $28525$ for the \cite{Kiefer2019} relation and of $9.03$ for the \cite{Karoff2009}--\cite{Metcalfe2007} relation. Their MAE to the combined data of both stages is 7.91 and 7.87, respectively. The combined MAE of the relations found here for these data is 4.28. For the main sequence individually, these values are 4.36 and 4.31 compared to 2.33 for the relation found here. For the subgiants and lower RGB stage they are 13.9 and 13.9 compared to 7.62 for the optimal relation. As can be seen, the relations from the literature do not reproduce the sensitivity factors as well as the empirical relations found here.


\begin{table*}
	\caption{Relations fitted to the theoretical mode sensitivities and their optimal parameters. Within each relation's paragraph, the six rows correspond to the six stages of evolution from Table~\ref{table:2} according to their letter symbol. The asterisks in the last column indicate the best model per evolutionary stage.}\label{table:3}
	\begin{tabular}{c|c|c|c|c|c|c|c|c|c|c}
		\hline
		Number &                                          Relation                                           & Stage&  $\alpha$ & $\beta$  & $\gamma$ & $\delta$ & $\epsilon$ & $\zeta$ &  MAE  & \\ \hline
		  1    &                                $ R^{\alpha}\cdot M^{\beta}$                                 &   A &    -3.56   &   14.7   &          &          &            &         & 8.86  &    \\
		       &                                                                                             &   B &    0.905   &   1.41   &          &          &            &         & 12.58 &    \\
		       &                                                                                             &   C &    0.624   &  -0.512  &          &          &            &         & 2.55  &    \\
		       &                                                                                             &   D &    0.138   &  -0.346  &          &          &            &         & 0.54  &    \\
		       &                                                                                             &   E &    0.518   &  -0.483  &          &          &            &         & 2.26  &    \\
		       &                                                                                             &   F &    -0.984  &   5.61   &          &          &            &         & 5.21  &    \\ \hline
		  2    &                  $ R^{\alpha}\cdot M^{\beta}\cdot T_{\text{eff}}^{\gamma}$                  &   A &    -1.08   &   4.31   &   0.28   &          &            &         & 2.50  &     \\
		       &                                                                                             &   B &    -4.79   &   8.71   &  0.471   &          &            &         & 8.22  &     \\
		       &                                                                                             &   C &    -0.582  &  0.234   &  0.346   &          &            &         & 2.38  &     \\
		       &                                                                                             &   D &    -0.847  & -0.00659 &  0.439   &          &            &         & 0.31  &     \\
		       &                                                                                             &   E &    -1.14   &  -0.189  &  0.503   &          &            &         & 2.31  &     \\
		       &                                                                                             &   F &    -1.56   &  -0.564  &  0.923   &          &            &         & 4.03  &     \\ \hline
		  3    &                        $ R^{\alpha}\cdot M^{\beta} \cdot L^{\delta}$                        &   A &    -0.348  &   30.2   &          &  -4.64   &            &         & 8.31  &       \\
		       &                                                                                             &   B &    -10.2   &  -9.83   &          &   7.13   &            &         & 10.43 &       \\
		       &                                                                                             &   C &     14.6   &   3.11   &          &  -8.99   &            &         & 2.29  & $\ast$ \\
		       &                                                                                             &   D &    -30.3   &  -7.34   &          &   19.6   &            &         & 0.36  &       \\
		       &                                                                                             &   E &    -8.33   &  -1.51   &          &   5.42   &            &         & 1.96  & $\ast$ \\
		       &                                                                                             &   F &    -11.2   &  -1.48   &          &   7.24   &            &         & 5.36  &     \\ \hline
		  4    &               $ R^{\alpha}\cdot M^{\beta} \cdot \nu_{\text{max}}^{\epsilon}$                &   A &    -0.554  &   4.29   &          &          &   0.301    &         & 2.53  &     \\
		       &                                                                                             &   B &    0.109   &  -0.198  &          &          &   0.319    &         & 8.25  &     \\
		       &                                                                                             &   C &    0.109   & -0.0856  &          &          &   0.369    &         & 2.38  &     \\
		       &                                                                                             &   D &    0.023   &  -0.419  &          &          &   0.476    &         & 0.31  &     \\
		       &                                                                                             &   E &    -4.92   &  -3.91   &          &          &    3.56    &         & 2.73  &     \\
		       &                                                                                             &   F &    0.381   &  -0.757  &          &          &   0.807    &         & 4.30  &     \\ \hline
		  5    & $ R^{\alpha}\cdot M^{\beta}\cdot T_{\text{eff}}^{\gamma} \cdot \nu_{\text{max}}^{\epsilon}$ &   A &    -13.4   &   5.5    &   6.61   &          &   -6.81    &         & 2.33  & $\ast$ \\
		       &                                                                                             &   B &    -13.7   &   4.67   &   7.12   &          &   -7.36    &         & 6.34  & $\ast$ \\
		       &                                                                                             &   C &    -13.0   &   6.26   &   6.53   &          &   -6.58    &         & 2.43  &      \\
		       &                                                                                             &   D &     5.08   &  -2.82   &  -2.55   &          &    3.24    &         & 0.31  &      \\
		       &                                                                                             &   E &     37.8   &  -25.5   &  -20.9   &          &    26.0    &         & 2.34  &      \\
		       &                                                                                             &   F &     12.2   &  -6.76   &   -5.9   &          &    7.29    &         & 4.29  &      \\ \hline
		  6    &       $ R^{\alpha}\cdot M^{\beta}\cdot L^{\delta} \cdot \nu_{\text{max}}^{\epsilon}$        &   A &    -4.23   &  -1.62   &          &   2.54   &   0.307    &         & 2.33  &       \\
		       &                                                                                             &   B &    -3.82   &   -3.0   &          &   2.74   &   0.315    &         & 6.34  &       \\
		       &                                                                                             &   C &    -37.3   &  -10.3   &          &   23.7   &   0.787    &         & 2.31  &       \\
		       &                                                                                             &   D &     1.54   & -0.0693  &          &  -0.98   &   0.495    &         & 0.31  &$\ast$ \\
		       &                                                                                             &   E &     8.85   &  -2.93   &          &  -8.05   &    3.44    &         & 2.34  &       \\
		       &                                                                                             &   F &     3.98   &  -0.406  &          &  -2.27   &    0.93    &         & 4.29  &       \\ \hline
		  7    &               $ T_{\text{eff}}^{\gamma} \cdot L^{\delta}\cdot \log g^{\zeta}$               &   A &            &          &  -3.28   &   1.84   &            &  20.7   & 2.43  &      \\
		       &                                                                                             &   B &            &          &  -3.18   &   1.79   &            &  20.4   & 7.62  &      \\
		       &                                                                                             &   C &            &          & 0.00889  & -0.00719 &            &   1.7   & 2.39  &      \\
		       &                                                                                             &   D &            &          &  0.317   &  -0.383  &            &  0.433  & 0.31  &      \\
		       &                                                                                             &   E &            &          &  0.603   &  -0.909  &            & 0.0101  & 2.36  &      \\
		       &                                                                                             &   F &            &          &  0.549   &  -0.457  &            &  0.87   & 4.20  & $\ast$ \\ \hline
	\end{tabular} 
\end{table*}

\begin{table*}
	\caption{Parameters of best fitting relations per evolutionary stage.}\label{table:4}
	\begin{tabular}{|c|c|c|c|c|c|c}
		\hline
		        Stage          &   $R$    &   $M$   & $ T_{\text{eff}}$ &   $L$    & $\nu_{\text{max}}$ & $\log g$ \\
		                       & $\alpha$ & $\beta$ &     $\gamma$      & $\delta$ &     $\epsilon$     & $\zeta$  \\ \hline
		    Main sequence      & -13.4  &  5.5  & 6.61 & &-6.81 & \\
		Subgiants \& lower RGB & -13.7  & 4.67 &  7.12 & &-7.36 &\\
		      Upper RGB        & 14.6 & 3.11 &&-8.99&&\\
		       Tip RGB         &  1.54 & -0.0693 &&-0.98  &0.495&\\
		   Core He burning     & -8.33 &-1.51 &&  5.42&& \\
		         AGB           & && 0.549 & -0.457 && 0.87\\ \hline
	\end{tabular}
\end{table*}

\subsection{Application of the scaling relations}\label{sec:51}
With the relations presented here and a measure of stellar activity, it is possible to estimate activity related frequency shifts for stars from the main sequence to the AGB. We note that the prediction of frequency shifts has to be done with a normalization to a known activity level, typically with the Sun as the standard. For example, one would take the average logarithmic fraction of the solar luminosity that is emitted in the Calcium II H and K lines $\log R'_{\text{HK}\odot}$, multiply it by the sensitivity value obtained from the main-sequence relation, and scale the result to obtain the known solar cycle p-mode frequency shift amplitude of \unit[0.4]{$\mu$Hz}. This scaling factor can then be applied to stellar $\log R'_{\text{HK}}$ values and the sensitivities from the presented relations to give the expected full-cycle frequency shift values of the respective stars. 

We use the main sequence' polynomial from Table~\ref{table:4} and known activity measures for the stars KIC~8006161 (HD~173701), KIC~10644235, and KIC~12009504 to predict their full-cycle frequency shifts. These stars' and the Sun's fundamental parameters as well as their $\log R'_{\text{HK}}$ and mean photometric activity $\langle S_{\text{ph}}\rangle$ are given in Table~\ref{table:5}. See \cite{Mathur2014a, Mathur2014} for the definition of $\langle S_{\text{ph}}\rangle$. The measured frequency shift amplitude $\delta\nu_{\text{meas}}$ was taken from \cite{Santos2018}. It is simply calculated as the difference between their reported maximum and minimum value of the p-mode frequencies.

For KIC~8006161 we find that using $\log R'_{\text{HK}}$ to predict the full-cycle shift gives $\delta\nu_{\text{pred,Ca}}=\unit[0.18\pm0.04]{\mu\text{Hz}}$, whereas the measured shift is $\delta\nu_{\text{meas}}=\unit[1.12\pm0.09]{\mu\text{Hz}}$. \cite{Karoff2013} already found that this star shows an unexpectedly low excess flux in the Calcium lines. Thus, $\log R'_{\text{HK}}$ appears to be an unsuitable activity measure to predict the p-mode shift of this star in particular. For KIC~10644253 the predicted shift using $\log R'_{\text{HK}}$ is much closer to the observed value and and the predicted and observed shifts are even in agreement, within the uncertainty, for KIC~12009504. It should be noted that the values of $\log R'_{\text{HK}}$ were not measured contemporaneously to the \textit{Kepler} data. Depending on at what phase of a star's activity cycle $\log R'_{\text{HK}}$ is measured, it potentially underestimates the full-cycle shifts.

Using the mean photometric activity index $\langle S_{\text{ph}}\rangle$ to predict the frequency shifts, we find  $\delta\nu_{\text{pred,Sph}}=\unit[0.71\pm0.15]{\mu\text{Hz}}$ for KIC~8006161,  \unit[0.91$\pm$0.24]{$\mu$Hz} for KIC~10644253, and  \unit[0.37$\pm$0.11]{$\mu$Hz} for KIC~12009504. These are in agreement with the measured values of KIC~10644253 and KIC~12009504. For KIC~8006161 this is again somewhat lower than the measured value. This can be due to two reasons: The sensitivity $S$ is underestimated or $\langle S_{\text{ph}}\rangle$ is underestimated. As can be seen from the top left panel in Figure~\ref{fig:5}, our best relation underestimates the sensitivity by a few tens of percent for stars with  $\nu_{\text{max}}$ around \unit[3500]{$\mu$Hz}. Furthermore, KIC~8006161 has an inclination of $38^{+4}_{-3}$ degrees \citep{Karoff2018}. This can lead to an underestimation of its photometric activity, as active regions that lead to a global p-mode frequency shifts, could hide on the out-of-sight portion of the star. Thus, it is likely that both factors contribute to the underestimation of this star's frequency shifts.

\subsection{Assumption of a global near-surface perturbation}\label{sec:52}
With Eq.~(\ref{sec3:eq3}) we located the cause of the frequency shifts entirely in the outer pressure scale height of the star with an isotropic distribution over the entire stellar sphere.

In a study of activity-related p-mode frequency shifts of \textit{Kepler} stars, \cite{Salabert2018a} found that for four stars of their 87 star sample, the shifts have an oscillatory behaviour as a function of mode frequency. Such a behaviour hints towards a cause of at least part of the frequency shifts which is located deeper within the resonant p-mode cavity, possibly at the depth of the He {\sc{II}} ionization zone. The sample of \cite{Salabert2018a} is limited both in quality and size but poses the very intriguing question of where the (magnetic) perturbation that is associated with the observed frequency shifts is actually concentrated in the radial direction. 

The frequency dependence of the shifts also depends on the spatial distribution of activity as was shown by, e.g., \cite{Kiefer2018}. Further, the spatial distribution of activity affects modes of different harmonic degree and azimuthal order differently \citep{Moreno2000}. Analysing the frequency shifts of the radial and dipole modes of KIC~8006161 (HD~173701), \cite{Thomas2019} found that the active regions for this star are at higher latitudes than in the solar case and are distributed over a wider range. 

We argue that any magnetic perturbation that is happening at a depth below the outer pressure scale height will eventually manifest in the near-surface layers. There it will then lead to frequency shifts that are larger than those caused by deeper seated perturbations \citep{Gough1990, Kiefer2018}. Further, we are not concerned with the frequency dependence of the frequency shifts here, but focus on the full-cycle shift of the radial mode closest to $\nu_{\text{max}}$. The assumption of the source function with Eq.~(\ref{sec3:eq3}) appears to be maintainable in this regard and certainly results in a very good estimate of the expected level of the mode sensitivities and thus of the frequency shifts. Still, given the results of \cite{Salabert2018a} and \cite{Thomas2019} we again stress the importance of more detailed investigations of frequency shifts as a function of mode frequency and of harmonic degree if the data allow for such analyses.

\begin{table*}
\caption{Fundamental parameters of test stars.}\label{table:5}
\centering
\begin{tabular}{|c|c|c|c|c|c|c|c|c|c}
	\hline
	\rule[-1ex]{0pt}{2.5ex} Star & $R$  & $M$ & $T_{\text{eff}}$ & $\nu_{\text{max}}$  & $\log R'_{\text{HK}}$ & $\langle S_{\text{ph}}\rangle$  & $\delta\nu_{\text{meas}}$  & $\delta\nu_{\text{pred,Ca}}$ &$\delta\nu_{\text{pred,Sph}}$ \\
	&[R$_{\odot}$]&[M$_{\odot}$] & [K]&$\left[\rm{\mu}\text{Hz}\right]$&&[ppm]&$\left[\rm{\mu}\text{Hz}\right]$&$\left[\rm{\mu}\text{Hz}\right]$&$\left[\rm{\mu}\text{Hz}\right]$\\
	\hline
	Sun &1&1&5777&3104&-4.901&166.1$\pm$2.6&0.4&&\\
	KIC~8006161 & 0.9293$\pm$0.0083 & 0.9861$\pm$0.0253 & 5488$\pm$77& 3574.7$\pm$11.4 & -5.03 & 492.4$\pm$6.3 & 1.12$\pm$0.09 &0.18$\pm$0.04&0.71$\pm$0.15\\
	KIC~10644253 & 1.1221$\pm$0.0133 & 1.1748$\pm$0.0385 & 6045$\pm$77 & 2899.7$\pm$22.8 & -4.689 & 385.1$\pm$8.9 & 1.07$\pm$0.25 &0.64$\pm$0.17&0.91$\pm$0.24\\
	KIC~12009504 & 1.4120$\pm$0.0193 & 1.1986$\pm$0.0452 & 6179$\pm$77 & 1865.6$\pm$7.7 & -4.949 & 131.2$\pm$3.4 & 0.49$\pm$0.16 &0.42$\pm$0.12&0.37$\pm$0.11\\
	\hline\\
\end{tabular}

{\raggedright \textbf{References.} Radius and mass: \cite{SilvaAguirre2017}. $T_{\text{eff}}$ and $\nu_{\text{max}}$: \cite{Lund2017} and references therein. $\log R'_{\text{HK}}$: \cite{Santos2019}, who used values from \cite{Karoff2013} and \cite{Salabert2016}. $\langle S_{\text{ph}}\rangle$: \cite{Mathur2014}; except KIC~8006161: \cite{Garcia2014}. Measured frequency shift $\delta\nu_{\text{meas}}$ from \cite{Santos2018}. Solar values of $\log R'_{\text{HK}}$ from \cite{Saar1999}, $\langle S_{\text{ph}}\rangle$ from \cite{Mathur2014}. \par}

\end{table*}

\section{Conclusions}\label{sec:6}
We computed the sensitivity of p-mode oscillations to magnetic activity for a set of evolutionary stellar models. For this, we used a standard MESA inlist and adjusted only the initial mass (from 0.7--3.0\,\unit{M$_{\odot}$}) and maximum time steps between models. We then used GYRE to compute p-mode eigenfunctions of all the models. The complete set of models was divided into six evolutionary stages. We found that modes are most sensitive to perturbations for main-sequence and subgiant stars. On the main sequence, sensitivity increases with higher mass. Lowest mode sensitivities are found for stars at the tip of the RGB.

In the calculation of the mode sensitivity $S$ we assumed that the leading terms in the kernel functions of p modes are proportional to $\text{div}\boldsymbol{\xi}$. We further assumed that the perturbations are concentrated within one pressure scale height below the photospheric radius of the models. Fits of polynomials in fundamental and widely available stellar parameters then gave relations for estimating these sensitivities.

In contrast to the relations for the mode sensitivity that have been used in the literature so far, when frequency shifts due to stellar activity were to be estimated, our relations are based on stellar models. The added flexibility of changing the exponents on the fundamental parameters in the polynomials between evolutionary stages certainly increases the capability of these relations to capture the true variation of the sensitivities through stellar evolution. The relation given by \cite{Karoff2009}, which is based on derivations of \cite{Metcalfe2007}, and the relation derived by \cite{Kiefer2019} are unable to reproduce the mode sensitivities of main-sequence stars and subgiant and lower RGB stars (which they were originally intended for) as well as the optimal relations presented here. In particular, the mass dependence of mode sensitivities and their change over time while the stars are on the main sequence is not well captured by these relations. For stars off the main sequence, these relations considerably overestimate mode sensitivities.

We used the optimal relation for the main-sequence and tested it on three stars. We found that it is important to use activity measures that are contemporaneous to the photometric data that are used to measure the p-modes whose shifts are to be predicted. If simple photometric activity indices are used, inclination of the stellar rotation axis can lead to an underestimation of the true level of activity and thus to underestimated shifts. Using the mean photometric activity index $\langle S_{\text{ph}}\rangle$, we find overall solid agreement between the predicted and observed frequency shifts for the three stars that we investigated.

The derivation of the mode sensitivities is done only for purely radial oscillations, which ensures that the modes are pure p modes. This is of particular importance in RGB stars, where, e.g., dipole modes, are often of mixed character, partly p mode and partly g mode. For non-radial modes, the calculation of the mode inertia and the kernel function would change, as the eigenfunctions include a horizontal part. It can be expected that mode sensitivities of mixed modes are weaker, as their mode inertia will have contributions from both the convective parts of the star (p mode) and the radiative part (g mode). 

The reported relations and the presented mode sensitivities can also guide the selection of targets for future studies of activity related p-mode frequency shifts: Mode sensitivities are largest for earlier main-sequence stars at around $\nu_{\text{max}}=\unit[2000]{\mu Hz}$. Thus, in turn, frequency shifts can be expected to be largest for these stars. Further, subgiants in the range $\unit[400]{\mu Hz}\lesssim\nu_{\text{max}}\lesssim\unit[800]{\mu Hz}$ show increased sensitivities and present good targets. Finally, we note that there appear several models with larger sensitivities in the upper RGB, core He burning, and the AGB stages. Such studies -- optimally including oscillating stars at various evolutionary stages -- could also aim at testing our relations, given known activity levels, fundamental stellar parameters, and p-mode frequency shifts. 

The cardinal question driving all of this is: how do stellar dynamos operate and in what way do they change along stellar evolution? Simple time resolved p-mode frequency measurements can play an important part in answering this.


\section*{Acknowledgements}
RK and A-MB acknowledge the support of the STFC consolidated grant ST/L000733/1. The authors thank the anonymous referee for taking the time to review this article and for their useful comments.

\textsc{software}: This work made use of the following software libraries not cited in the text: \textsc{matplotlib} \citep{2007CSE.....9...90H}, \textsc{mesaPlot} \citep{robert_farmer_2020_3678482}, \textsc{MESASDK} \citep{richard_townsend_2019_3560834}, \textsc{NumPy} \citep{Oliphant2006}, \textsc{pandas} \citep{Pandas2019}, \textsc{SciPy} \citep{2020SciPy-NMeth}, \textsc{seaborn} \citep{Seaborn2018}, \textsc{Uncertainties}, \citep{uncertainties}.

\section*{Data Availability}
The datasets underlying this article were generated with publicly available software and can be reproduced using the details provided in the appendix:
\textsc{MESA}, \href{http://mesa.sourceforge.net/}{http://mesa.sourceforge.net/},
\textsc{GYRE}, \href{https://bitbucket.org/rhdtownsend/gyre/}{https://bitbucket.org/rhdtownsend/gyre/}

The data underlying this article will be shared on reasonable request to the corresponding author.


\bibliographystyle{mnras}


\appendix

\section{Inlists}\label{sec:app:1}
\subsection{MESA inlists}\label{sec:app:1.1}
The following inlists were used to generate the set of stellar models with MESA. As described in Section~\ref{sec:2}, only the values for \path{initial_mass} and \path{max_years_for_timestep} were changed. As can be seen from these inlists, the maximal time between the calculated stellar models is ten times smaller than those stated in Table~\ref{table:1} but we only saved every tenth model (\path{profile_interval = 10}). All keys which have the default MESA value are marked with a \#, which should be removed before this inlist is used.

First, this inlist is used to evolve the star to the main sequence:
\begin{verbatim}
&star_job
    show_log_description_at_start = .false.
    create_pre_main_sequence_model = .true.
    save_model_when_terminate = .true.
    save_model_filename = 'start.mod'
    required_termination_code_string = 
        'max_model_number'
#    kappa_file_prefix = 'gs98'
    pre_ms_relax_num_steps = 50
    pgstar_flag = .true.
/ ! end of star_job namelist

&controls
    max_years_for_timestep= 5d7
    initial_mass = 0.7 
    
#    use_gold_tolerances = .true.
#    use_eosELM = .true.
#    use_eosDT2 = .true.
    
    max_number_backups = 1
    max_number_retries = 5
    max_model_number = 50
    
#    initial_z = 0.02d0
    use_Type2_opacities = .true.
    Zbase = 0.02d0
      
    varcontrol_target = 1d-3
    
    photo_interval = 50
    write_model_with_profile = .true.
    write_pulse_data_with_profile = .true.
    pulse_data_format = 'GYRE'

    profile_interval = 10
    model_data_prefix = 'model' 
    model_data_suffix = '.mod'
    max_num_profile_models = 1500

    history_interval = 10
    terminal_interval = 10
    write_header_frequency = 10
/ ! end of controls namelist
\end{verbatim}

After this, the star is evolved until the end of the AGB with the following inlist:

\begin{verbatim}
&star_job
    show_log_description_at_start = .false.
    
    load_saved_model = .true.
    saved_model_name = 'start.mod'
    save_model_when_terminate = .true.
    save_model_filename = 'end_agb.mod'
    required_termination_code_string 
        = 'envelope_mass_limit'
    
#    kappa_file_prefix = 'gs98'
    
    change_initial_net = .true.
    new_net_name = 'o18_and_ne22.net'
    
    new_surface_rotation_v = 2 ! solar (km sec^1)
    set_near_zams_surface_rotation_v_steps = 10 
    
    change_D_omega_flag = .true.
    new_D_omega_flag = .true.
    
    set_initial_cumulative_energy_error = .true.
    new_cumulative_energy_error = 0d0
    
    set_initial_age = .true.
    initial_age = 0
    
    set_initial_model_number = .true.
    initial_model_number = 0
    
    pgstar_flag = .true.
/ ! end of star_job namelist
    
&controls
    max_years_for_timestep= 5d7
    initial_mass = 0.7
    
#    use_gold_tolerances = .true.
#    use_eosELM = .true.
#    use_eosDT2 = .true.
    
    use_eps_mdot = .true.
    
    use_dedt_form_of_energy_eqn = .true.
    min_cell_energy_fraction_for_dedt_form = 0
    use_eps_correction_for_KE_plus_PE_in_dLdm_eqn 
        = .true.
    
    num_trace_history_values = 2
    trace_history_value_name(1) = 'rel_E_err'
    trace_history_value_name(2) 
        = 'log_rel_run_E_err'
    
    max_number_backups = 10
    max_number_retries = 160
    max_model_number = 6500
    backup_hold = 3
    
#    initial_z = 0.02d0
    
    use_Type2_opacities = .true.
    Zbase = 0.02d0
    
    am_nu_visc_factor = 0
    am_D_mix_factor = 0.0333333333333333d0
#    D_DSI_factor = 0
    D_SH_factor = 1
    D_SSI_factor = 1
    D_ES_factor = 1
    D_GSF_factor = 1
    D_ST_factor = 1
    
    varcontrol_target = 1d-3
    delta_lgL_He_limit = 0.01d0
    
    envelope_mass_limit = 1d-2 ! Msun
    
    cool_wind_full_on_T = 9.99d9
    hot_wind_full_on_T = 1d10 
    cool_wind_RGB_scheme = 'Reimers'
    cool_wind_AGB_scheme = 'Blocker'
#    RGB_to_AGB_wind_switch = 1d-4
    Reimers_scaling_factor = 0.8d0  
    Blocker_scaling_factor = 0.7d0
    
    photo_interval = 50
    write_model_with_profile = .true.
    write_pulse_data_with_profile = .true.
    pulse_data_format = 'GYRE'
    profile_interval = 10
    model_data_prefix = 'model' 
    model_data_suffix = '.mod'
    max_num_profile_models = 1500
    history_interval = 10
    terminal_interval = 10
    write_header_frequency = 10
/ ! end of controls namelist
\end{verbatim}

\subsection{GYRE inlist}\label{sec:app:1.2}
Name lists that were not used (e.g, \path{'&nad_output'}) are not listed here. The frequency interval to be scanned by GYRE, set by the \path{freq_min} and \path{freq_max} keys, is determined as $\left[\nu_{\text{max}}-10\cdot\Delta\nu,\nu_{\text{max}}+10\cdot\Delta\nu\right]$, with a minimum value for \path{freq_min} of \unit[0.001]{$\mu$Hz}. Here, $\nu_{\text{max}}$ is calculated as described in Section~\ref{sec:3} and the large frequency separation $\Delta\nu$ is calculated for each model via the scaling relation of \cite{Kjeldsen1995}:
\begin{align}
    \Delta\nu = \Delta\nu_{\odot}\left(\frac{M}{M_{\odot}}\right)^{0.5}\left(\frac{R}{R_{\odot}}\right)^{-1.5},\label{a1:eq1}
\end{align}
with $\Delta\nu_{\odot}=\unit[135]{\mu Hz}$.
\begin{verbatim}
&model
    model_type = 'EVOL'
    file = 'path_to_MESA_gyre_file'
    file_format = 'MESA'
/
    
&mode
    l = 0
/
    
&osc
    outer_bound = 'JCD'
    inertia_norm = 'RADIAL'
/
    
&num
    diff_scheme = 'COLLOC_GL4'
/
    
&scan
    grid_type = 'LINEAR'
    freq_min = freq_min_value
    freq_max = freq_max_value
    freq_min_units = 'UHZ'
    freq_max_units = 'UHZ'
    n_freq = 500
/

&grid
    alpha_osc = 20
    alpha_exp = 4
    n_inner = 10
/

&ad_output
    summary_file = 'path_to_summary_file'
    summary_file_format = 'TXT'
    summary_item_list = 'M_star,R_star,l,n_pg,
        freq,freq_units,E,E_p,E_g,E_norm'
    mode_template = 'path_of_output_file'
    mode_file_format = 'TXT'
    mode_item_list = 'l,n_pg,freq, freq_units,E,
    E_p,E_g,E_norm,x,xi_r,xi_h,rho,P'
    freq_units = 'UHZ'
/
\end{verbatim}

\bsp	
\label{lastpage}
\end{document}